\documentclass{aa}
\usepackage{hyperref}
\usepackage{multirow}
\usepackage{graphicx}
\usepackage{amsmath}	
\usepackage{amssymb}	

\usepackage{txfonts}
\usepackage{natbib} 
\usepackage[usenames]{color}

\newcommand{\obj}{RX\, J1131$-$1231}

\bibpunct{(}{)}{;}{a}{}{,}
\begin{document}

\title{ALMA view of \obj: Sub-kpc CO (2-1) mapping of a molecular disk in a lensed star-forming quasar host galaxy}
\author{D. Paraficz\inst{1} 
\and M. Rybak\inst{2,3} 
\and J. P. McKean\inst{4,5}
\and S. Vegetti\inst{3} 
\and D. Sluse\inst{6} 
\and F. Courbin\inst{1}
\and H. R. Stacey\inst{4,5} 
\and S. H. Suyu\inst{3,8,9}
\and M. Dessauges-Zavadsky\inst{10}
\and C. D. Fassnacht\inst{7} 
\and L. V. E. Koopmans\inst{4} 
}

\institute{Institute of Physics, Laboratory of Astrophysics,
Ecole Polytechnique F\'ed\'erale de Lausanne (EPFL)
Observatoire de Sauverny
CH-1290 Versoix, Switzerland \label{EPFL}
\and
Leiden Observatory, Leiden University, PO Box 9513, NL-2300 RA Leiden, the Netherlands
\and
Max Planck Institute for Astrophysics, Karl-Schwarzschild-Strasse 1, 85740 Garching, Germany
\and
Kapteyn Astronomical Institute, University of Groningen, P.O. Box 800, 9700 AV Groningen, The Netherlands
\and
ASTRON, Netherlands Institute for Radio Astronomy, Postbus 2, NL-7990 AA, Dwingeloo, the Netherlands
\and
STAR Institute, Quartier Agora - All\'ee du six Ao\^ut, 19c B-4000 Li\`ege, Belgium
\and
Department of Physics, University of California, Davis, CA 95616, USA
\and 
Physik-Department, Technische Universit{\"a}t M{\"u}nchen, James-Franck-Strasse 1, D-85748 Garching, Germany
\and 
Institute of Astronomy and Astrophysics, Academia Sinica, PO Box 23-141, Taipei 10617, Taiwan
\and
Observatoire de Gen\`eve, Universit\'e de Gen\`eve, 51 Chemin des Maillettes, 1290 Versoix, Switzerland
}
\date{\today}
\titlerunning{Resolved ALMA view of \obj}
\authorrunning{D. Paraficz}

\abstract{We present ALMA 2-mm continuum and CO (2-1) spectral line imaging of the gravitationally lensed $z=0.654$ star-forming/quasar composite \obj\ at 240 to 400~mas angular resolution. The continuum emission is found to be compact and coincident with the optical emission, whereas the molecular gas forms a complete Einstein ring, which shows strong differential magnification. 
The de-lensed source structure is determined on 400-parsec-scales resolution using a Bayesian pixelated visibility-fitting lens modelling technique.
The reconstructed molecular gas velocity-field is consistent with a large rotating disk with a major-axis FWHM $\sim$9.4~kpc at an inclination angle of $i=54^{\circ}$ and with a maximum rotational velocity of $280$~km\,s$^{-1}$. From dynamical model fitting we find an enclosed mass within 5 kpc of $M(r<5~{\rm kpc})=(1.46\pm0.31) \times 10^{11}$~M$_\odot$.
The molecular gas distribution is highly structured, with clumps that are co-incident with higher gas velocity dispersion regions (40--50 km\,s$^{-1}$) and with the intensity peaks in the optical emission, which are associated with sites of on-going turbulent star-formation. The peak in the CO (2-1) distribution is not co-incident with the AGN, where there is a paucity of molecular gas emission, possibly due to radiative feedback from the central engine. The intrinsic molecular gas luminosity is $L'_{CO}= 1.2\pm0.3\times10^{10}$~K~km\,s$^{-1}$~pc$^2$ and the inferred gas mass is $M_{H_2} = 8.3\pm3.0 \times 10^{10}$~M$_{\odot}$, which given the dynamical mass of the system is consistent with a CO--H$_2$ conversion factor of $\alpha = 5.5\pm2.0$~M$_{\odot}$\,(K~km\,s$^{-1}$~pc$^2$)$^{-1}$. This suggests that the star-formation efficiency is dependent on the host galaxy morphology as opposed to the nature of the AGN. The far-infrared continuum spectral energy distribution shows evidence for heated dust, equivalent to an obscured star-formation rate of ${\rm SFR} = 69^{+41}_{-25} \times (7.3 / \mu_{\rm IR}) $~M$_{\odot}$~yr$^{-1}$, which demonstrates the composite star-forming and AGN nature of this system. 
}


\maketitle
\section{Introduction}

Constraining the kinematic and morphological properties of galaxies, tracing their interstellar medium (ISM) and establishing their star formation properties, requires high angular-resolution observations that can trace molecular gas on scales of a few tens to around a hundred parsecs. With its long-baseline capability (up to 16 km) the Atacama Large Millimetre/submillimetre Array (ALMA) now provides cutting-edge observations of star forming galaxies and active galactic nuclei (AGN) on unprecedented spatial scales \citep{2015ApJ...808L...4A}. Yet, such studies are still limited to objects in the local Universe (e.g. \citealt{2014A&A...565A..97C}). For example, at $z=0.5$, details on 10-100 parsec-scales require an angular resolution of 1.5-15~mas, which is even beyond the capabilities of ALMA.

Using strong gravitational lensing as a natural telescope helps to circumvent this issue. Increasing both the apparent flux-density and solid-angle, gravitational lensing makes it possible to spatially resolve galaxies that would otherwise be impossible to observe with current facilities, or alternatively, allows the use of more compact interferometric arrays that provide a better sampling of the $uv$-plane and require less complex calibration procedures. In this paper, we take advantage of both the enhanced spatial resolution provided by ALMA and of the magnifying power of strong gravitational lensing to study the distribution and kinematics of the molecular gas in the host galaxy of \obj , discovered by \citet{2003A&A...406L..43S}.

\begin{figure*}
\begin{center}
\includegraphics[width=18cm,angle=0]{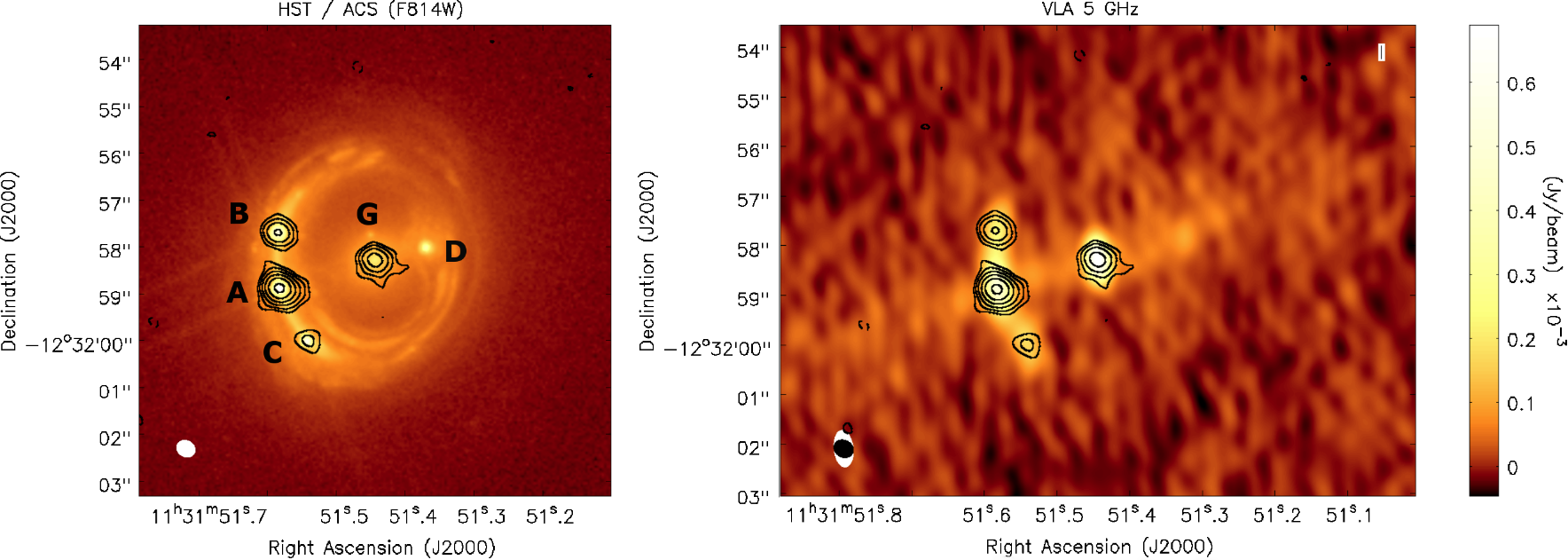}
\caption{(Left) HST/ACS image of \obj\ in the F814W filter taken from  \citet{2014ApJ...788L..35S}. (Right) VLA image at 5 GHz. In each panel, the ALMA 2.1 mm continuum contours at the (-3, 3, 6, 12, 24, 48, $96)\times10~\mu$Jy~beam$^{-1}$ level are overlaid for reference. The synthesised beams are presented in the lower-left corner of each panel.} 
\label{fig:hst-vla}
\end{center}
\end{figure*}

\obj\ is an exceptional strongly lensed quasar. It consists of four images of a $z_q = 0.654$ quasar (Fig.~\ref{fig:hst-vla}) with a total magnification of $\sim$50 at optical wavelengths \citep{2007A&A...468..885S}. The background quasar is hosted by a luminous late-type galaxy that is lensed by a foreground early-type galaxy at $z_l$ = 0.295 \citep{2003A&A...406L..43S,2007A&A...468..885S}. The host galaxy of the quasar is seen as a prominent Einstein ring in optical and infrared imaging with the {\it Hubble Space Telescope} ({\it HST}) (Fig.~\ref{fig:hst-vla}). 

As a consequence, \obj\ is one of the best studied lensed quasars found so far, with a wealth of observations over a broad range of wavelengths at many epochs. Previous observations include {\it Chandra} X-ray monitoring for a micro-lensing analysis \citep{2006A&A...449..539S, 2007A&A...468..885S, 2010ApJ...709..278D, 2012ApJ...757..137C}; integral field spectroscopy (IFS) for the characterization of low-mass substructure within the lensing halo  \citep{2007ApJ...660.1016S}; and an X-ray observation to estimate the accretion disk spin \citep{2014Natur.507..207R}. In addition, \obj\ is being photometrically monitored bi-weekly in the optical band by {\textsc COSMOGRAIL} \citep{2013A&A...556A..22T} in order to improve the gravitational lensing time-delay measurement between the different quasar images. The {\textsc COSMOGRAIL} monitoring and time delays, together with detailed gravitational lens mass modelling based on {\it HST} imaging and the stellar kinematics of the lens \citep{2014ApJ...788L..35S}, made \obj\ one of the most accurate cosmological probes available so far aside from cosmic microwave background, baryonic acoustic oscillations and standard candles \citep{2013ApJ...766...70S, 2017MNRAS.468.2590S, 2016JCAP...08..020B, 2016MNRAS.462.3457C, 2017MNRAS.465.4914B}.

In this paper, we have added another dimension to the datasets available for \obj\, by taking ALMA continuum and spectral-line imaging in the extended configuration (out to 1.6~km baselines). The data, obtained in Band 4 (145 GHz), were taken to spatially resolve the CO (2-1) emission line of the lensed host galaxy. With these new observations, we mapped the line intensity and velocity-field across the source, finding evidence for an Einstein ring. We then reconstructed the de-lensed source surface brightness distribution from the visibility data (this paper) allowing us to investigate in detail the AGN host kinematics and to further refine the gravitational lens mass model (Rizzo et al., in prep). Indeed, regions of different velocities in the source are affected by lensing in different ways as they lie at different spatial positions with respect to the centre of the lensing mass. Each velocity channel therefore provides an independent lensing configuration, for a fixed lensing potential. In a companion paper (Sluse et al. 2017), we use our ALMA data in the continuum to measure the flux-ratios between the quasar lensed images of \obj\ and compare these with the flux-ratios at other wavelengths. In Sluse et al. (2017), we show for the first time that micro-lensing is likely occurring at mm-wavelengths, and this allows us to place constraints on the size and origin of the mm-wavelength continuum emission region of the AGN.

Our paper is arranged as follows. Section \ref{obs} describes the observations, data reduction and data analysis. Section \ref{results} presents the observed image-plane properties of the data at mm, radio and optical wavelengths. Section \ref{modelling} describes the mass modelling and the source reconstruction. 
Section \ref{recons} discusses the intrinsic properties of the molecular gas. Finally, in Section~\ref{concl}, we discuss our results and present our conclusions. We use a flat $\Lambda$CDM cosmology with H$_0=69.6~{\rm km\,s}^{-1}~{\rm Mpc}^{-1}$, $\Omega_m=0.286$ and $\Omega_{\Lambda}=0.714$ \citep{2016A...594A..13P}.

\section{Observations}
\label{obs}
In this section, we present the new ALMA and {\it Herschel} observations of \obj\, and archival imaging with the VLA, that we have used for our analysis.

\subsection{Atacama Large Millimetre Array}

\obj\ was observed with ALMA on 2015 July 19 (Proposal Code: 2013.1.01207.S; PI: D. Paraficz), using 37 out of 54 of the 12~m array antennas. The purpose of the observations was to spatially resolve the redshifted CO (2-1) emission line of the background source, which has a rest-frequency of 230.538 GHz. The data were taken in Band 4 through four spectral windows. Three spectral windows were used in the standard continuum mode, each with 128 spectral channels and 2 GHz bandwidth tuned to central observing frequencies of 137.169, 149.118 and 150.993 GHz. One spectral window was used in spectral line mode, centred on the redshifted CO (2-1) line at 139.044 GHz, with 480 spectral channels and a total bandwidth of 1.875 GHz. This provided a spectral resolution of 8.4~km\,s$^{-1}$~channel$^{-1}$. The data were taken using both linear polarisations (XX and YY). The telescope configuration had projected baselines between 27.5 m and 1.6 km, which provided a sensitivity to structures with a largest angular size of about 16.2\arcsec. 

Titan and Ganymede were used to determine the absolute flux-density calibration, and the unresolved quasars J1058+0133 and J1118$-$1232 were used for calibrating the bandpass and correcting for the residual delays, respectively. The observations were phased-referenced using J1130$-$1449, with a cycle time between the target and calibrator of around 4.5 to 6 mins. The total observing time was about 2.5 h, with about 1.3 h on-source. The data were reduced and calibrated using the standard ALMA pipeline using the Common Astronomy Software Application package ({\sc casa}; \citealt{2007ASPC..376..127M}). The corrected visibilities for the target and calibrators that were produced by the pipeline reduction were found to have no major calibration errors. Self-calibration was not used as there was an insufficient signal-to-noise ratio on the continuum emission to provide good calibration solutions.

The imaging and de-convolution of the continuum and spectral line data were also carried out within {\sc casa} using the standard {\sc clean} method -- we note that the gravitational lens modelling of the data is carried out directly in the visibility plane and the imaging presented in Section~\ref{results} is to illustrate the observed image-plane properties of the data. The continuum imaging used all four spectral windows, but without those channels that contained the emission line, which resulted in a central frequency of 144.081~GHz (2.08~mm). Images were made using a Briggs weighting scheme with a robust parameter of 0 and 2 (2 being equivalent to natural weighting), which resulted in a beam size of $0.28\arcsec \times 0.24\arcsec$ and $0.41\arcsec \times 0.34\arcsec$, respectively, and an rms map noise of 15 and 10~$\mu$Jy~beam$^{-1}$, respectively.

The spectral line cube was formed using natural weighting and all 480 spectral channels within the spectral window containing the emission line. These data were cleaned using a mask and a detection threshold of $3-\sigma$ to remove any strong side-lobe emission, while limiting any contribution from noise peaks within the cube. The resulting beam size is $0.44\arcsec \times 0.36\arcsec$ with an rms map noise of about 340~$\mu$Jy~beam$^{-1}$ for a 8.4~km\,s$^{-1}$ channel width. From the cleaned spectral line cube, a velocity integrated map (moment-zero) of the line emission was made using those channels covering the line, intensity weighted velocity (moment-one) and velocity dispersion (moment-two) maps were made using those pixels with a signal-to-noise ratio $>3$.

\subsection{Herschel Space Observatory}

\obj\ was observed at far-infrared wavelengths using the \textit{Herschel Space Observatory} on 2012 December 08 (Proposal Code: OT1$\_$abercian$\_$1; PI: Berciana-Alba) as part of a study of over 100 lensed quasars \citep{stacey17}. The target was observed at 250, 350 and 500 $\mu$m for 90-s using the Spectral and Photometric Imaging Receiver (SPIRE) in small-map mode. The data were retrieved from the \textit{Herschel} archive and processed within the Herschel Interactive Processing Environment (HIPE; version 14.2.1). The flux density of the target was measured at the three far-infrared wavelengths using the time-line extractor.
\begin{figure}[!ht]
\begin{center}
\includegraphics[width = 8cm,angle=0]{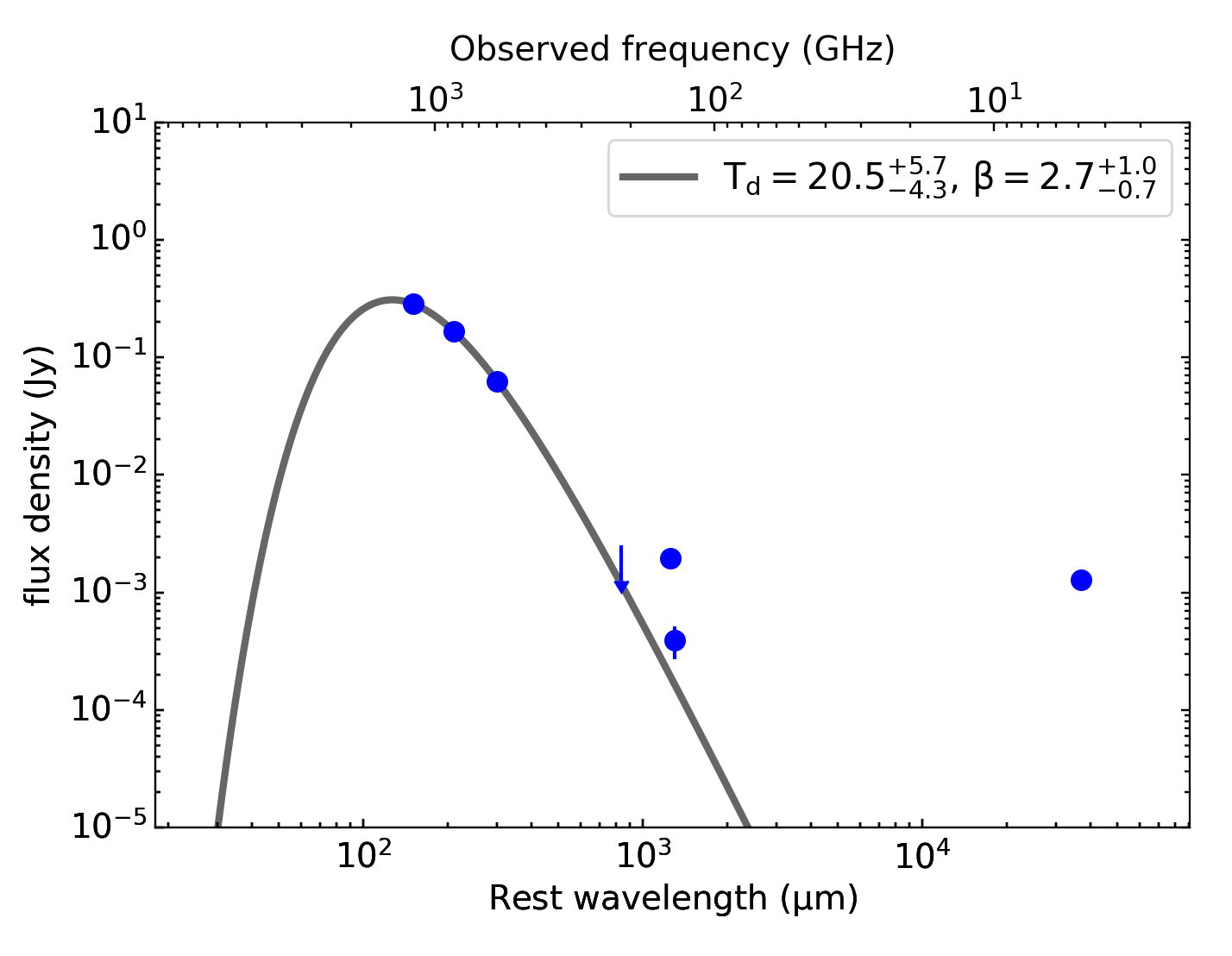}
\caption{Observed spectral energy distribution of \obj\ with the multi-wavelength photometry of the lensed images, not corrected for the gravitational lensing magnification (blue points with error bars). From left to right: {\it Herschel} 250, 350 and 500 $\mu$m (this work), CARMA 1.4~mm continuum \citep{2017ApJ...836..180L}, ALMA 2.1~mm continuum (this work), PdBI 2.2 mm continuum \citep{2017ApJ...836..180L} and the VLA 4.86 GHz (this work). The rest-frame best-fit modified black body spectrum for the heated dust is shown in grey, with an effective dust temperature $T=20.5^{+5.7}_{-4.3}$ K and dust emissivity index $\beta=2.7^{+1.0}_{-0.7}$.}
\label{fig:SED}
\end{center}
\end{figure}
\subsection{Very Large Array}

\obj\ was observed with the VLA at 4.86 GHz on 2008 December 29 (Proposal Code: AW741; PI: Wucknitz). The VLA was used in A-configuration, which provided baselines between about 0.5 and 33.5 km. In total, 25 out of 27 VLA antennas were available. The observations were phased referenced using J1130$-$148, which was used to determine the complex gains (amplitude and phase) as a function of time, and 3C286 was used to determine the absolute flux-density scale. The visibility integration time was 10~s and two spectral windows (intermediate frequencies) with 50 MHz bandwidth were used.

The data were calibrated in the standard way within {\sc casa}. Given the expected brightness of the target, and the likelihood that the side-lobes from field sources would limit the dynamic range, wide-field imaging using a natural weighting scheme of the visibilities was used. This resulted in a beam size of $0.74\arcsec \times 0.38\arcsec$ and an rms map noise of 14~$\mu$Jy~beam$^{-1}$.\\

\begin{table*}
\footnotesize
\begin{center}
\caption{ALMA mm-continuum positions, peak surface brightness and flux-densities for the lensed quasar images (A, B, C and D) and for the lensing galaxy (G), together with their relative positions in the optical \citep{2006A&A...451..865C, 2013ApJ...766...70S}. The absolute positions are quoted relative to the phase-referenced position of image A ($\rm RA = 1$1:31:51.582, $\rm Dec = -1$2:31:58.88). The uncertainty in the surface brightness and flux-density is assumed to be 10\%.}
\label{imagefluxes}
\begin{tabular}{rrrrrrrrrrr}
  \hline \hline

ID&RA& Dec&RA&Dec&RA& Dec& $I^{\rm peak}$ & $S_{\rm 2.1~mm}$&$\mu_{\rm AGN}$ \\

 & \multicolumn{2}{c}{This work}& \multicolumn{2}{c}{\citet{2006A&A...451..865C}} & \multicolumn{2}{c}{\citet{2013ApJ...766...70S}}& Jy~beam$^{-1}$ & Jy\\
 \hline
 \\
A& =0&=0&=0&=0&=0&=0&$1.1\times10^{-3}$& $1.4\times10^{-3}$&$-$21.29 \\
B& $+0.029\pm0.006$&$+1.196\pm0.006$& $+0.032\pm0.002$ &$+1.188\pm0.002$&0.039&1.182&$2.6\times10^{-4}$ &$3.3\times10^{-4}$&14.43 \\
C& $-0.615\pm0.018$&$-1.113\pm0.012$&$-0.590\pm0.003$ &$-1.120\pm0.001$&$-$0.557&$-$1.112&$7.5\times10^{-5}$& $9.9\times10^{-5}$&11.29 \\
D&$\cdots$ &$\cdots$&$-3.112\pm0.003$& $+0.884\pm0.002$ &3.111&$-$0.876& $<3.0\times10^{-5}$ & $<3.0\times10^{-5}$&$-$1.187 \\
G& $-2.035\pm0.007$&$+0.604\pm0.007$&$-2.016\pm0.002$ &$+0.610\pm0.002$&$-$2.028&$-$0.599&$3.4\times10^{-4}$& $4.9\times10^{-4}$& $\cdots$   \\

 \hline   
\end{tabular}
\end{center}
\end{table*}

\section{Observed image-plane properties}
\label{results}

In this section, we discuss the image-plane properties of the ALMA continuum and CO (2-1) emission, and compare with the image-plane data from the {\it HST} and the VLA.

\subsection{Continuum emission}

In Fig.~\ref{fig:hst-vla}, we present the continuum emission detected from \obj\ at 2.1~mm. We find from our ALMA imaging that there are four compact components associated with the gravitational lens; three are coincident with the optical/infrared point-source emission detected from lensed images A, B and C, and one component is associated with the lensing galaxy. We find no evidence of any continuum emission associated with lensed image D down to a $3-\sigma$ detection limit of $<30~\mu$Jy~beam$^{-1}$. The combined flux-density of the three lensed images A, B and C is $S_{\rm 2.1~mm} = 1.95\pm0.20$~mJy, and their flux-ratios seem to be inconsistent with the expectations for a typical cusp configuration, where the sum of the flux density from images B and C should be equivalent to the flux density of image A (see Sluse et al. 2017 for a discussion of the likely micro-lensing origin of this flux-ratio anomaly). The emission from the lensing galaxy is $S_{\rm 2.1~mm} = 0.49\pm0.05$~mJy.

Recent 2.2~mm observations of \obj\ by \citet{2017ApJ...836..180L} using the Plateau de Bure Interferometer (PdBI) at $\sim4\times 2$~arcsec angular resolution, also found evidence for extended continuum emission from the gravitational lens system, which they correctly attribute to be from both the lensing galaxy (motivated by the previous detection of radio emission from the lensing galaxy with the VLA and MERLIN; \citealt{2008evn..confE.102W}; see also Fig.~\ref{fig:hst-vla}) and the lensed images; here our ALMA imaging at 0.3 arcsec angular resolution (Briggs weighting; Robust = 0) clearly separates the individual components. The radio-mm spectral index of the lensing galaxy emission is $\alpha^{\rm 6~cm}_{\rm2~mm} = -0.21$, which is consistent with a flat-spectrum radio source (this assumes a power-law spectrum; $S_{\nu} \propto \nu^{\alpha}$). This, coupled with the extended jet emission seen from the VLA at 5 GHz confirms the initial interpretation by \citet{2017ApJ...836..180L} that there is excess mm-emission, and this is due to AGN emission from within the lensing galaxy. This is the third case of a foreground lensing galaxy having emission at mm-wavelengths, the others being SDP.81 \citep{2015ApJ...808L...4A} and the 8 o'clock arc (McKean at al, in prep.), which may have implications for interpreting low-resolution data at far-infrared to mm-wavelengths from ALMA and {\it Herschel} for other lens systems.

In Fig.~\ref{fig:SED}, we present the observed spectral energy distribution (SED) from \obj, excluding the emission from the lensing galaxy; here we expect the contribution at shorter wavelengths to be negligible given the spectral index of the emission and radiative cooling from synchrotron emission. We find that the far-infrared emission is consistent with a modified black body, with a cold dust temperature of $T_{\rm dust} = 21.0^{+6.0}_{-4.6}$~K and dust emissivity of $\beta = 2.6^{+1.0}_{-0.7}$. These values are consistent with typical dusty star-forming galaxies (e.g. \citealt{2010MNRAS.409L..13C}). However, we find that from the lensed images, there is also an excess of emission at 2~mm that could be either due to free-free emission from H\,{\sc ii} regions associated with star-formation, or possibly from synchrotron emission from the lensed quasar.

Finally, we note that there is a significant difference in the emission seen at radio and mm-wavelengths, where the VLA (see Fig.~\ref{fig:hst-vla}) and MERLIN \citep{2008evn..confE.102W} imaging shows an extended arc, whereas the mm-emission has three compact components. The radio emission that forms the arc between images A, B and C is thought to be due to star-formation processes as it is resolved out at mas-scale angular resolution with very long baseline interferometry (VLBI) (Olaf Wucknitz, priv. comm.). Further, sensitive imaging with e-MERLIN and VLBI can confirm this. Nevertheless, the excess of mm-emission and the different morphology suggests that the ALMA continuum emission is not probing the heated dust associated with obscured star-formation.

\subsection{CO (2-1) emission}
\label{COemission}
\begin{figure}[!h]
\begin{center}
\includegraphics[scale=0.38,angle=0]{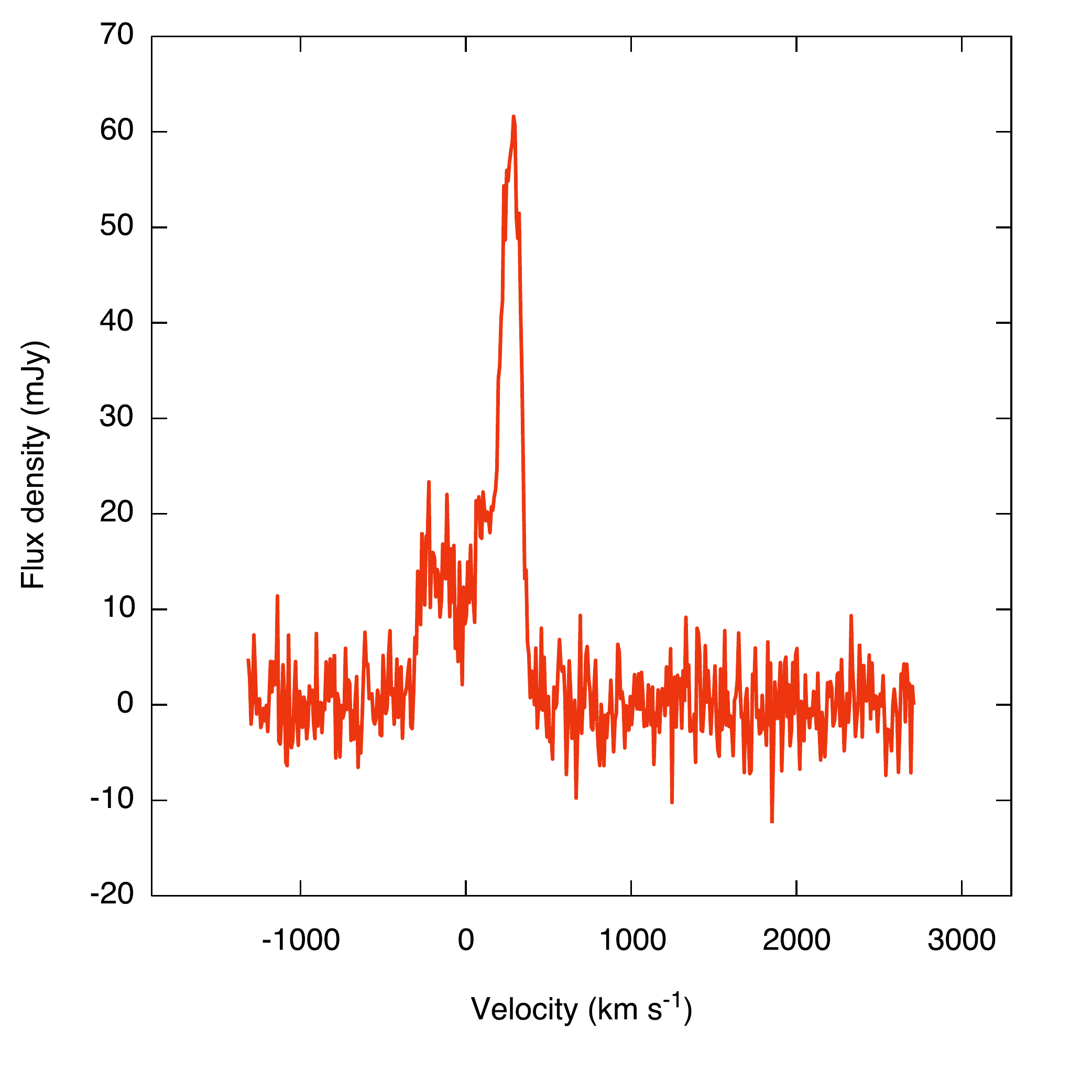}
\caption{Spatially integrated CO (2-1) emission line profile for \obj. The asymmetric double-horned profile is due to differential magnification across the source. The systemic velocity corresponds to a redshift of $z = 0.654$, and is the LSRK (Local Standard of rest - Kinematic) frame using the radio definition of the redshift.}
\label{fig:1d_line}
\end{center}
\end{figure}

\begin{figure*}[!ht]
\begin{center}
\includegraphics[scale=0.42,angle=0]{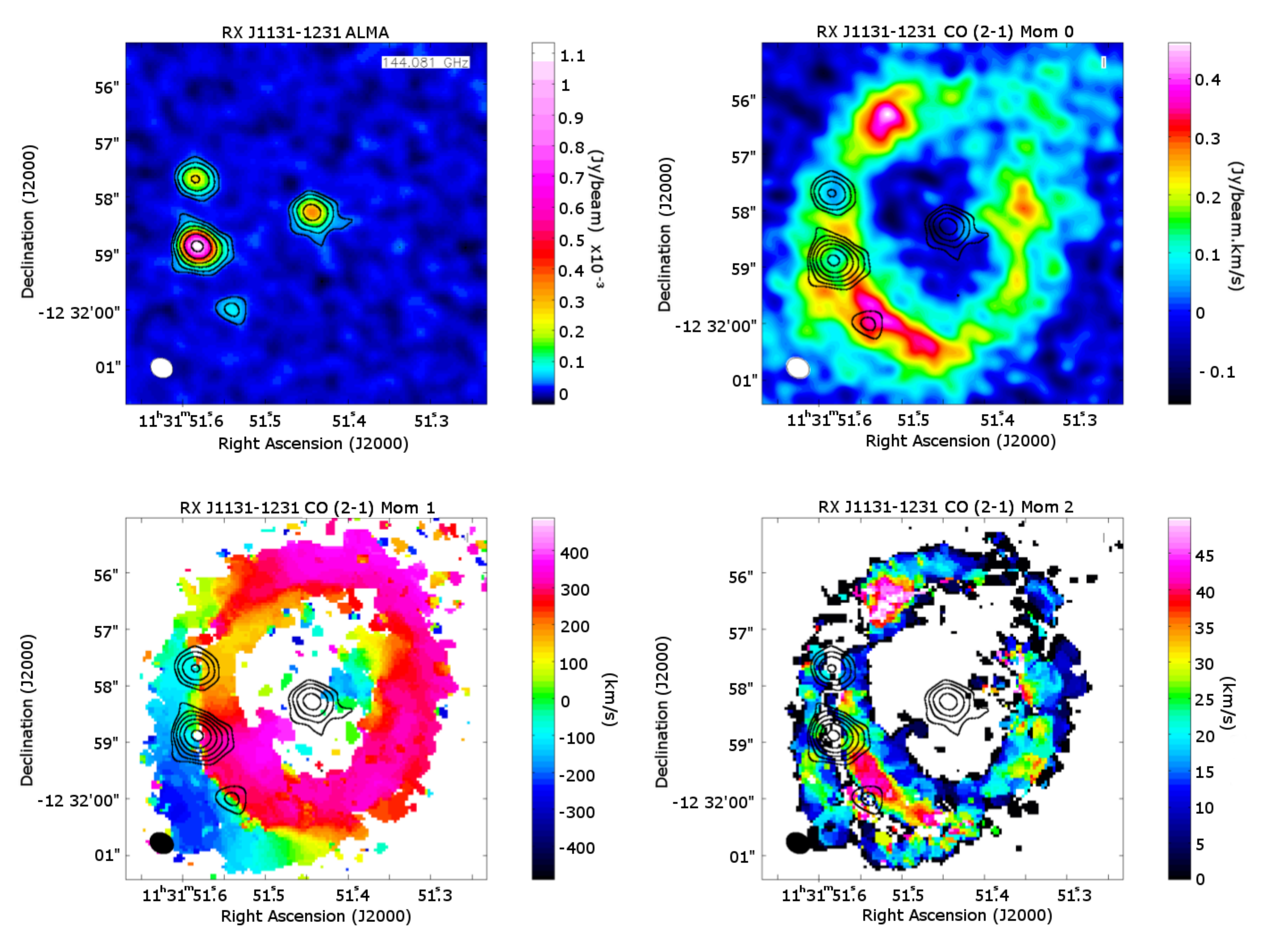}
\caption{(Upper-left) Naturally weighted continuum emission from \obj\ at 2.1~mm. Continuum emission from three of the lensed images and at the position of the lensing galaxy has been detected. The beam size is $0.41\arcsec \times 0.34$\arcsec at a position angle of 62 degrees east of north, and the rms noise level is 10~$\mu$Jy~beam$^{-1}$. (Upper-right) Naturally weighted CO (2-1) integrated line emission (moment-zero) map for \obj, showing an Einstein ring structure. The beam size is $0.44\arcsec \times 0.36 \arcsec$ at a position angle of 60~degrees east of north, and the rms noise level is 0.038~Jy~beam$^{-1}$~km\,s$^{-1}$. (Lower-left) Naturally weighted CO (2-1) intensity weighted velocity (moment-one) map for \obj, showing the clear velocity dependent structure of the molecular Einstein ring emission. (Lower-right) Naturally weighted CO (2-1) intensity weighted velocity dispersion (moment-two) map for \obj, showing evidence for turbulent molecular gas regions with a dispersion between 10 to 50~km\,s$^{-1}$. The beam size of the moment-one and moment-two maps are the same as the moment-zero map. In each panel, the ALMA 2.1 mm continuum contours at the (-3, 3, 6, 12, 24, 48, 96)$\times10~\mu$Jy~beam$^{-1}$ level are overlaid for reference.}
\label{fig:alma-moments}
\end{center}
\end{figure*}

\begin{figure*}
\begin{center}
\includegraphics[scale=0.55,angle=0]{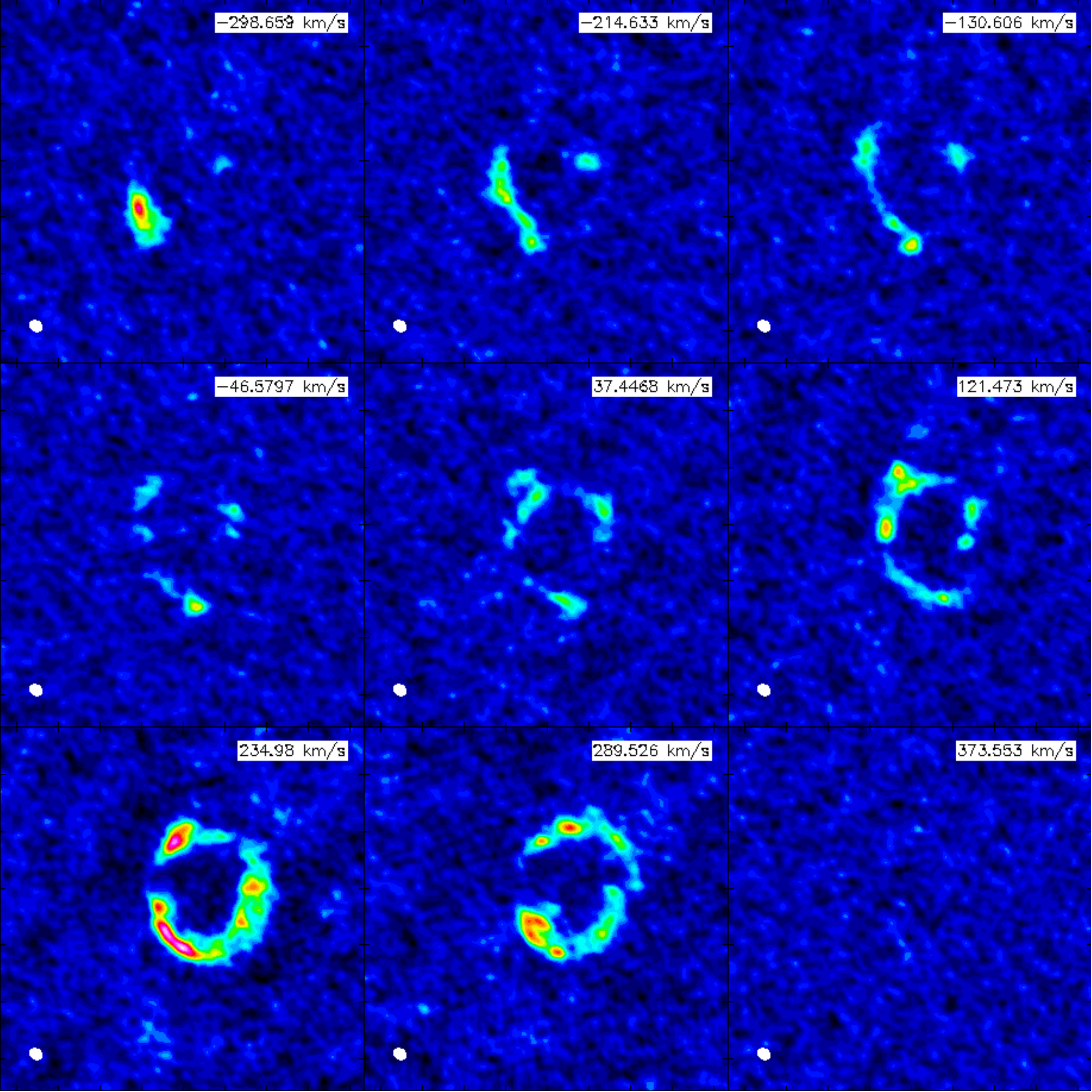}
\caption{Channel maps (82 km\,s$^{-1}$ width) for the CO (2-1) emission line from about $-300$ to $+400$~km\,s$^{-1}$, relative to the systemic velocity. The maps show the clear velocity dependent structure of the line surface brightness distribution. Each map covers a sky area of $12.8 \arcsec \times12.8\arcsec$ and the linear intensity-scale ranges from $-0.04$ to 0.2~Jy~beam$^{-1}$~km\,s$^{-1}$. The systemic velocity corresponds to a redshift of $z = 0.654$, and is in the LSRK frame using the radio definition of the redshift. The synthesised beam has a size of $0.44\times0.36$ \arcsec at a position angle of 60~degrees east of north, and is shown in the bottom left-hand corner of each map as the white ellipse.}
\label{fig:alma-mos}
\end{center}
\end{figure*}

In Fig.~\ref{fig:1d_line}, we show the integrated CO (2-1) spectrum over the full extent of \obj, and in Fig.~\ref{fig:alma-moments} we show the moment-zero, -one and -two maps. The integrated spectrum shows a double horn profile, which is typically associated with a rotating disk, that is also highly asymmetric. \citet{2017ApJ...836..180L} also found an asymmetric emission profile in their PdBI CO (2-1) and CARMA CO (3-2) spectra, albeit at a much lower signal-to-noise ratio. In Fig.~\ref{fig:alma-mos}, we show image slices of the CO (2-1) line, which clearly demonstrates the position dependent structure of the molecular gas emission as a function of velocity. Similar to the case of SDP.81 \citep{2015ApJ...808L...4A}, the asymmetric line profile can be attributed to differential magnification (e.g. \citealt{Rybak2015a}; see also discussion by \citealt{2017ApJ...836..180L}). This highlights the importance of carrying out a proper reconstruction of the surface brightness distribution of the background object, when it is extended, as a single value for the magnification is no longer valid.

The velocity integrated line intensity map (moment-zero) shows the CO (2-1) forms a complete Einstein ring of emission which total is well detected at the 76-$\sigma$ level, and has a highly structured surface brightness distribution. We also find no evidence for CO (2-1) emission at the position of the lensing galaxy, which further supports that the continuum emission at that location is from the lensing galaxy, as opposed to a core-lensed image. 

The velocity field shows a spectacular Einstein ring in velocity space that is dominated, due to differential magnification, by the redshifted gas component. There is also evidence for variations in the velocity dispersion from around 10 to 50 km\,s$^{-1}$ that are correlated with the intensity changes in the line emission. From Fig.~\ref{fig:alma-moments}, it is clear that the peaks in the velocity integrated molecular gas distribution and the regions with the largest velocity dispersion are not coincident with the quasar emission. We will discuss these correlations further in Section~\ref{recons}.

\section{Lens modelling and source reconstruction}
\label{modelling}
The distorting effect of the foreground lensing galaxy on the background object is corrected for using the visibility-fitting lens modelling technique (see \citealt{Rybak2015b,Rybak2015a}), which is an extension to the visibility domain of the Bayesian pixellated technique developed by \citet{2009MNRAS.392..945V}. 

In short, for a given observational noise $\boldsymbol{n}$ (assumed to be Gaussian and uncorrelated), this method relates the observed complex visibility function $\boldsymbol{d}$ and the unknown background source surface brightness distribution $\mathbf{}{s}$ by a set of linear equations,
\begin{equation}
\mathbf{FL}(\psi(\boldsymbol{\eta},\boldsymbol{x}))\boldsymbol{s} + \mathbf{n} = \boldsymbol{d}.
\end{equation}
Here, $\mathbf{L}$ is the lensing operator that relates the un-lensed surface brightness distribution, $\boldsymbol{s}$, to a model lensed surface brightness distribution and $\mathbf{F}$ is the response operator, which relates the model image-plane surface brightness distribution $\mathbf{L}\boldsymbol{s}$ to a model visibility function. $\psi(\boldsymbol{\eta},\boldsymbol{x})$ denotes the lensing potential for a given set of lens-model parameters $\boldsymbol{\eta}$ and $\boldsymbol{x}$ the image-plane position. By construction, $\mathbf{F}$ contains information about the sampling function of the interferometer and the primary beam profile of the individual antennas. The un-lensed surface brightness distribution $\boldsymbol{s}$ is reconstructed on an adaptive Delaunay tessellation with magnification-dependent resolution (see \citealt{2009MNRAS.392..945V} for details). 

The projected mass density profile of each lensing galaxy is described by an elliptical power-law. Specifically, our mass model has eight free parameters: the normalised surface-mass density, ellipticity, position angle, centre coordinates, mass-density slope as a function of radius (only for the main galaxy; we assume the satellite to follow an isothermal mass profile), and the external shear strength and position angle. We refer to these parameters collectively as $\boldsymbol{\eta}$. 

The optimisation is performed in the visibility space via the following penalty function,
\begin{equation}
P(\mathbf{s},\boldsymbol{\eta},\lambda~|~\boldsymbol{d})\propto\left[\mathbf{FL}\boldsymbol{s}-\boldsymbol{d}\right]^T\mathbf{C_d}^{-1} \left[\mathbf{FL}\boldsymbol{s}-\boldsymbol{d}\right] + \lambda^2 ~\|\mathbf{H} \boldsymbol{s}\|^2_2,
\end{equation}
where $\mathbf{C_d}$ is the diagonal covariance matrix of the data and $\lambda$ is the regularisation constant that encodes the level of smoothness of the background source surface brightness distribution, which is also a free parameter of the model. 

\begin{figure*}
\begin{center}
\includegraphics[scale=0.67,angle=0]{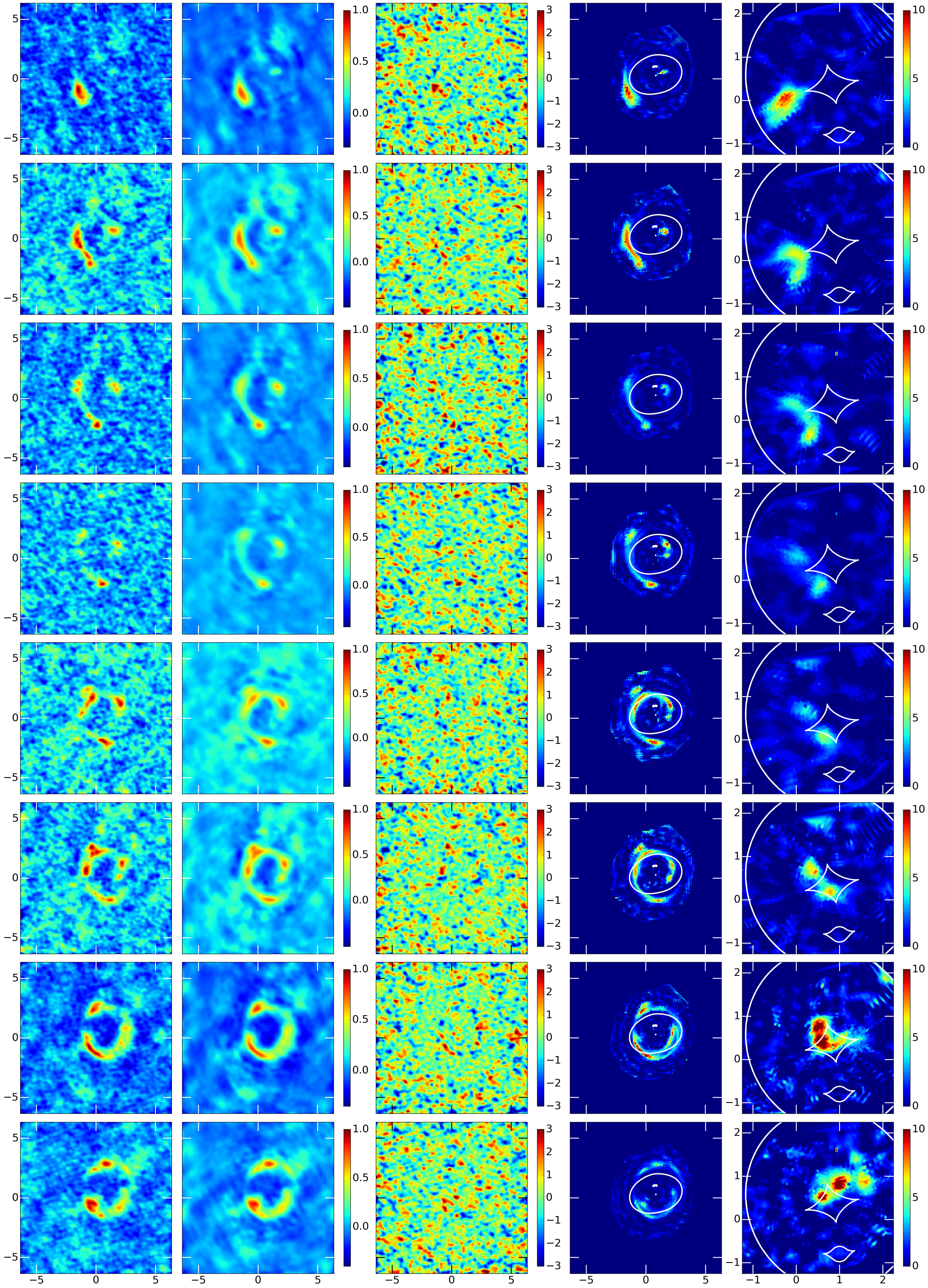}
\end{center}
\caption{Reconstruction of the CO (2-1) emission for our 8 velocity channels. From left to right are shown the dirty image of the data, the dirty image of the best model (normalised to the peak of the dirty image data), the residuals of the data$-$model, the best image-plane model and the best source-plane model (in units of mJy~km\,s$^{-1}$~arcsec$^{2}$). 
The white lines denote the critical lines and caustics in the image plane and the source plane, respectively. The axis coordinates are in arcsec with respect to the ALMA phase-tracking centre. The spatial scale in the source plane (right panels) is $1\arcsec = 7.030$~kpc.}
\label{fig:source_CO}
\end{figure*}

We find the best fitting mass distribution parameters $\boldsymbol{\eta}_{best}$ by combining all channels in the CO (2-1) emission line together to improve the overall signal-to-noise ratio and to provide the largest number of observational constraints; recall that the observed structure at different velocities will probe different parts of the lensing potential. We then split the data into eight velocity channels and remodel them individually by keeping the mass model fixed to $\boldsymbol{\eta}_{best}$ and by deriving the most likely {\it a posteriori} source surface brightness distribution and regularisation level for each individual velocity channel of width 84~km\,s$^{-1}$. The reconstructions of the 8 velocity channels studied here are presented in Fig.~\ref{fig:source_CO}, which shows a good agreement between the observed and model image plane data.

We find that the inferred lensing mass model parameters (see Table \ref{tab:lens_model}) are in agreement generally within 2-$\sigma$ with those derived by \citet{2013ApJ...766...70S} and \citet{2016JCAP...08..020B} from a modelling of the {\it HST} data, as well as with those derived by \citet{2016MNRAS.462.3457C} from a modelling of Keck-II adaptive optics data at 2.2~$\mu$m (with the largest discrepancy being on the mass density slope of the main galaxy). Both the {\it HST} and Keck-II data have an angular resolution that is over a factor of 5 better than the ALMA Band 4 observations presented here; the coarser resolution of the ALMA data shows itself in the large scatter in the derived parameters for the satellite galaxy's mass distribution.

In contrast, our model is not consistent with the lens parameters recently derived by \citet{2017ApJ...836..180L}, who used imaging with the PdBI of the CO (2-1) line at $\sim4\times 2$~arcsec angular resolution, which also shows tensions with \citet{2013ApJ...766...70S}, \citet{2016JCAP...08..020B} and \citet{2016MNRAS.462.3457C}. In particular, \citet{2017ApJ...836..180L} do not account for the presence of the satellite galaxy to the north of the main lensing galaxy, and do not allow for an external shear contribution; this leads to a mass model that is significantly flattened ($q=0.56$), and could bias their CO (2-1) emission line reconstruction.

\begin{table*}[!t]
\caption{Parameters of the best lens model for \obj, derived from the integrated CO (2-1) data. The position of the main lens is given with respect to the ALMA phase-tracking centre; the position of the satellite is given with respect to the centre of the main lens. \label{tab:lens_model}}
\begin{center}
 \begin{tabular}{@{}c|cccccccc @{}}
 \hline \hline
& $\kappa_0$ & $q$ & $\theta$ & $\Delta X$ & $\Delta Y$ & $\gamma$ & $\Gamma$ & $\Gamma_\theta$ \\
& [arcsec] & & [deg] & [arcsec] & [arcsec] & & & [deg]\\ 
 \hline
Main Lens & 1.70$\pm$0.09 & 0.767$\pm$0.014 & 111$\pm$1 & $-$0.83$\pm$0.01 & 0.28$\pm$0.14 & 2.04$\pm$0.03 & 0.084$\pm$0.016 & 93$\pm$3 \\
Satellite & 0.19$\pm$0.05 & 0.890$\pm$0.081& 1$\pm$22 & $-$0.07$\pm$0.10 & 0.74$\pm$0.14 & $\equiv$ 2 & $\equiv$ 0 & $\equiv$ 0 \\
 \hline
 \end{tabular}
\end{center}
\end{table*}

\begin{figure*}
\begin{center}
\includegraphics[scale=0.45,angle=0]{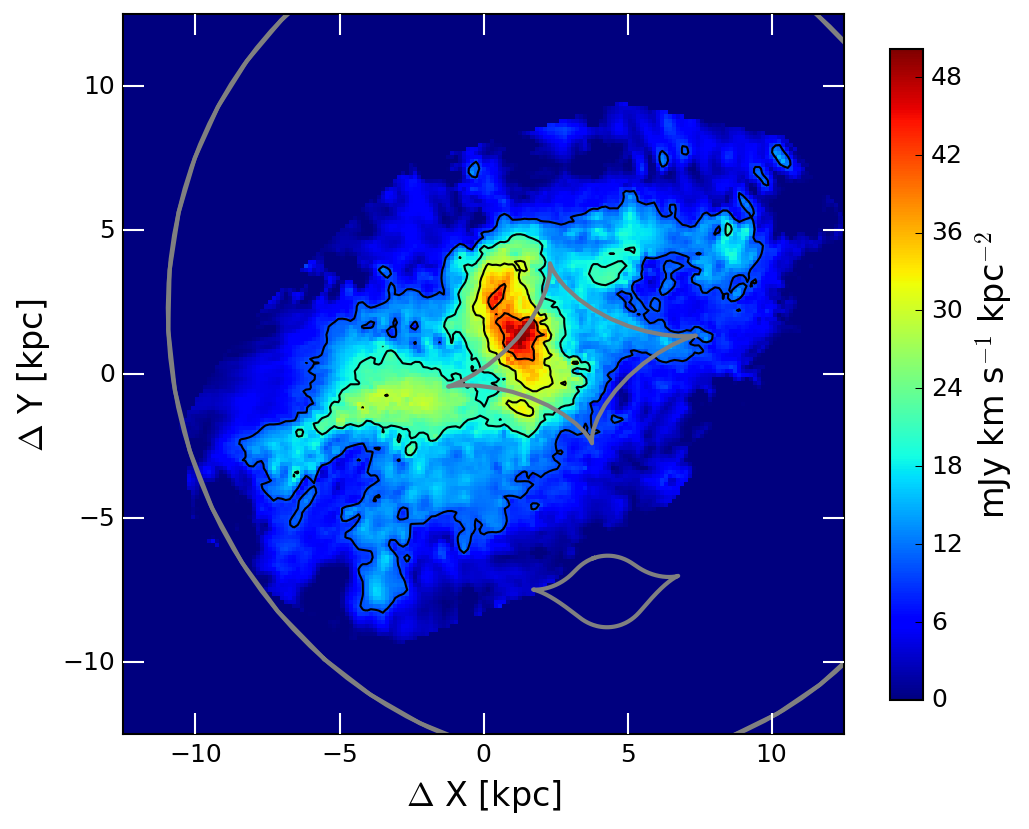}
\includegraphics[scale=0.45,angle=0]{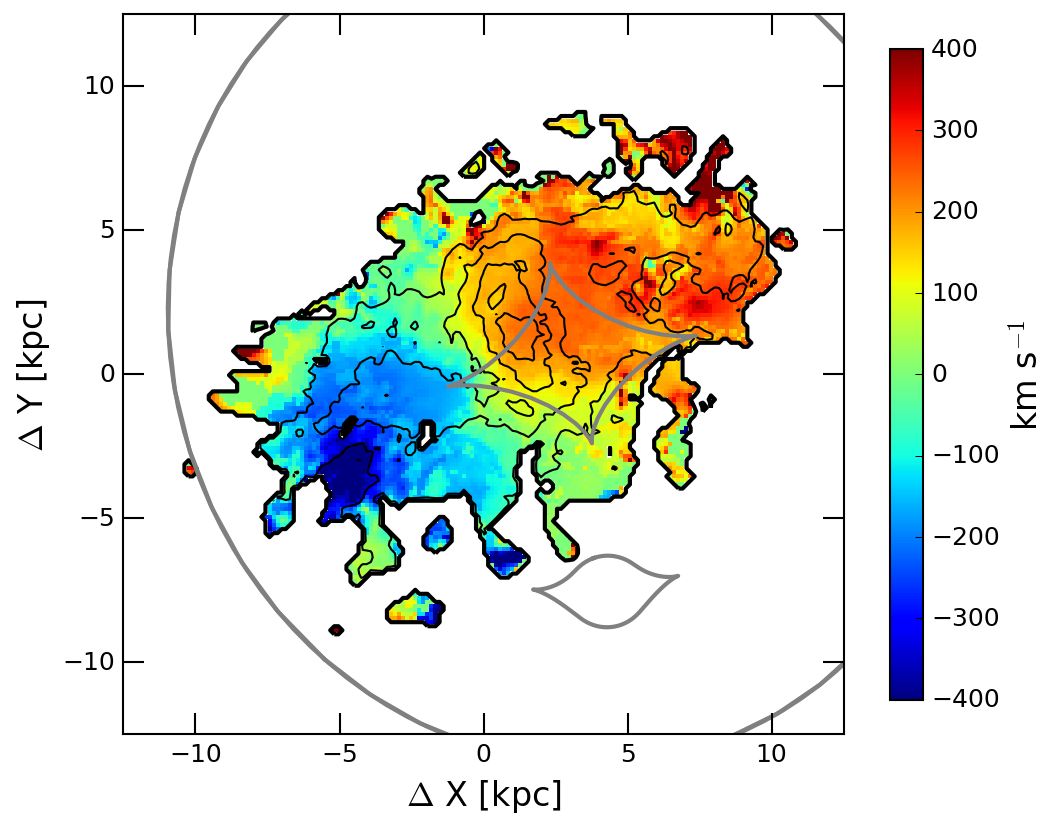}
\hspace*{-12mm}
\includegraphics[scale=0.45,angle=0]{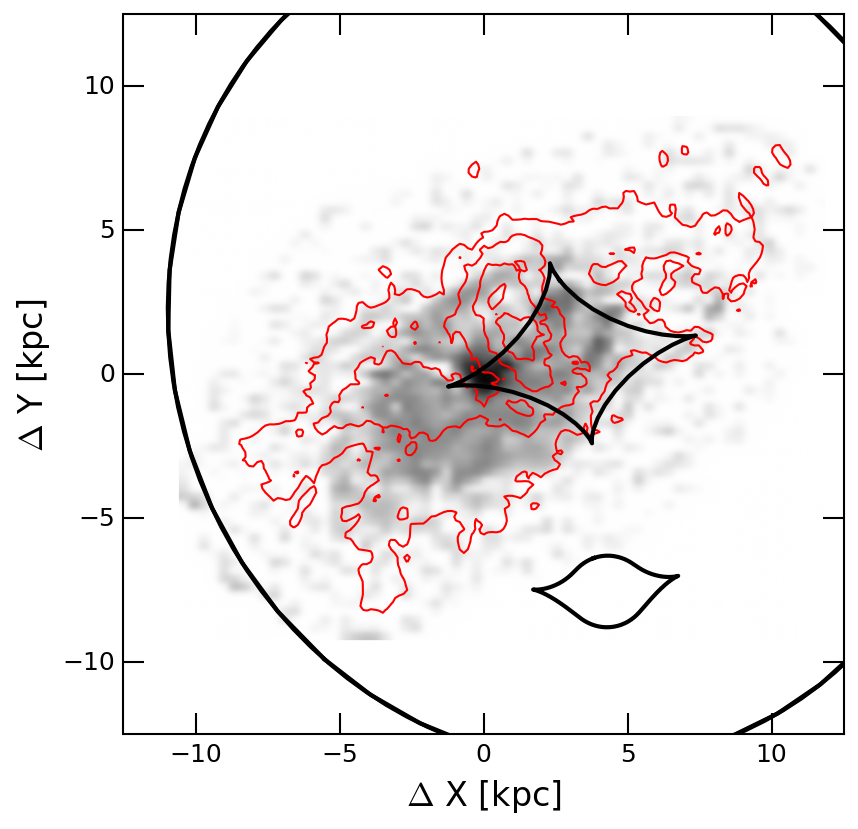}
\includegraphics[scale=0.4,angle=0]{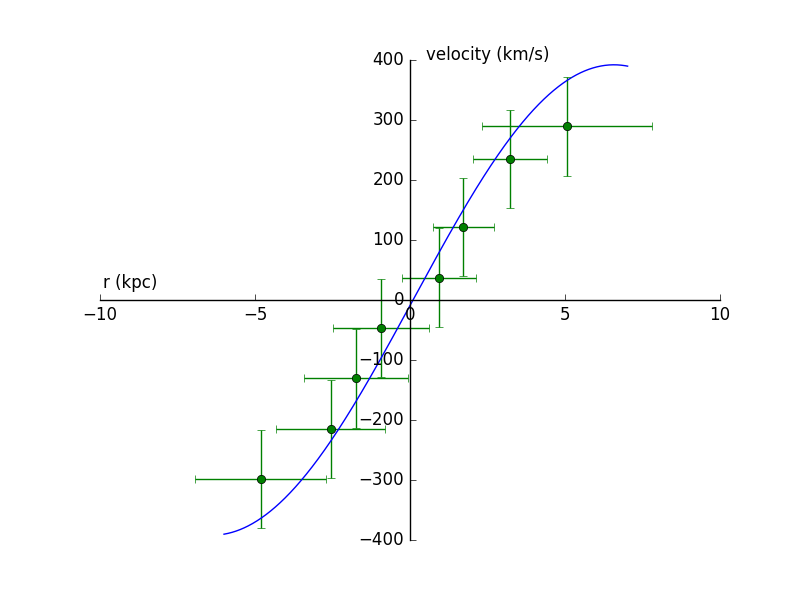}
\caption{(Upper-left) Moment-zero (integrated line intensity, in units of mJy~km\,s$^{-1}$~kpc$^{-2}$). (Upper-right) Moment-one (velocity field, in units of km\,s$^{-1}$). For the moment-one map, only regions with a signal-to-noise ratio $> 1$ are taken into account. The thin black lines trace the 25, 50 and 75 per cent contours of the moment-zero peak intensity. The coordinate system is centred on the AGN position obtained from the HST imaging \citep{2013ApJ...766...70S}. (Lower-left) Overlay of the reconstructed CO (2-1) moment-zero map (red contours) on the optical continuum image (grey scale, \citealt{2013ApJ...766...70S}).
The source-plane caustics are shown as solid grey lines. (Lower-right) One-dimensional rotation curve of the reconstructed CO (2-1) source, displaying a shape typical for disk-like dynamics. The blue solid line represents the rotation curve, corrected for disk inclination of $i=54^{\circ}$.} 
\label{fig:CO_recontructed_all}
\end{center}
\end{figure*}

\section{Intrinsic source properties}
\label{recons}

In this section, we discuss the reconstructed source to infer the intrinsic properties of the host galaxy of \obj.

\subsection{CO (2-1) intensity and velocity distribution}

In Fig.~\ref{fig:CO_recontructed_all}, we present the reconstructed integrated CO (2-1) line intensity and velocity field maps, and the molecular gas distribution relative to the reconstructed optical emission from the host galaxy. From both the molecular gas and optical emission, we see the clear disk like structure of the host galaxy, which is distributed over the full extent of the source ($\sim21$~kpc). The major-axis FWHM of the CO (2-1) surface brightness distribution is 9.4$\pm$1.0~kpc. The velocity field shows the typical structure that we would expect to see for a rotating disk. 

We find that the CO (2-1) emission has a complex distribution that is clumpy, showing a two spiral arm structure. There is also clear evidence that the peak in the CO (2-1) emission is not coincident with the centre of the host galaxy, and hence, the location of the AGN, but may be part of a bar that is close to the centre of the host galaxy. From comparing with Fig.~\ref{fig:alma-moments}, we see that this peak in the CO (2-1) distribution is also coincident with the region of higher velocity dispersion, demonstrating that this region has increased gas turbulence. It is not clear if the enhanced brightness distribution and velocity dispersion in the CO molecular gas in this region of the source is associated with a site of ongoing star-formation or if the offset from the central engine is due to feedback from the AGN. We note that the velocity field in this region is consistent with the overall smooth structure of the rotating disk, with no evidence for outflowing gas. Therefore, it is most likely that the peak in the CO (2-1) distribution is associated with star-formation. Further observations of, for example, the mm-continuum at a higher frequency closer to the peak in the dust bump of the spectral energy distribution may show evidence of dust obscured star-formation in this region.

From our reconstructed CO (2-1) intensity map, we derive an intrinsic line intensity of $I_{\rm CO}=2.06\pm0.43$~Jy km\,s$^{-1}$, which corresponds to a line luminosity of $L'_{CO}= 1.2\pm0.3\times10^{10}$~K~km\,s$^{-1}$~pc$^2$ (see Solomon at al 1997). We find that our calculation of the intrinsic line intensity is about 30\% lower than that found by \citet{2017ApJ...836..180L}, most likely due to the differences in the lens models and the fact that the PdBI dataset was taken at a lower angular resolution. In particular, we associate this discrepancy with the mass model differences, as our mass model predicts a factor of 2 higher magnification in the red channels, which dominate the image-plane line intensity. Moreover, \citet{2017ApJ...836..180L} detected CO emission as far out as $\sim$400~km\,s$^{-1}$, which we do not see in our deeper ALMA images; this adds an additional $\sim$10\% to their inferred $I_{\rm CO}$.

Finally, although \citet{2017ApJ...836..180L} discuss a possible CO (2-1) component from a source-plane companion satellite galaxy in their PdBI data. However, we find no evidence for this satellite component in our higher resolution and higher sensitivity spectral line imaging with ALMA.

\subsection{Star-formation rate, far-infrared luminosity and dust mass} 

We have determined the intrinsic properties of the dust emission from the modified black body spectrum that was fitted to the far-infrared and mm-wavelength continuum emission (see Fig,~\ref{fig:SED}). This fit is limited by the precision of differential magnification and the lack of short wavelength data that would properly define the cold dust temperature and the possibility that there is also heated dust emission from the AGN at short wavelengths. For these calculations, we assume that the dust emission is traced by the molecular gas distribution and hence has the same magnification ($\mu_{CO} = 7.3$). We find a total infrared luminosity of $L_{\rm 8-1000~\mu m} = 4.14^{+2.56}_{-1.50} \times (7.3 / \mu_{\rm IR}) \times 10^{11}$~L$_{\odot}$ and a star-formation rate of ${\rm SFR} = 69^{+41}_{-25} \times (7.3 / \mu_{\rm IR}) $~M$_{\odot}$~yr$^{-1}$.

\begin{figure*}
\begin{center}
\includegraphics[height=6cm,angle=0]{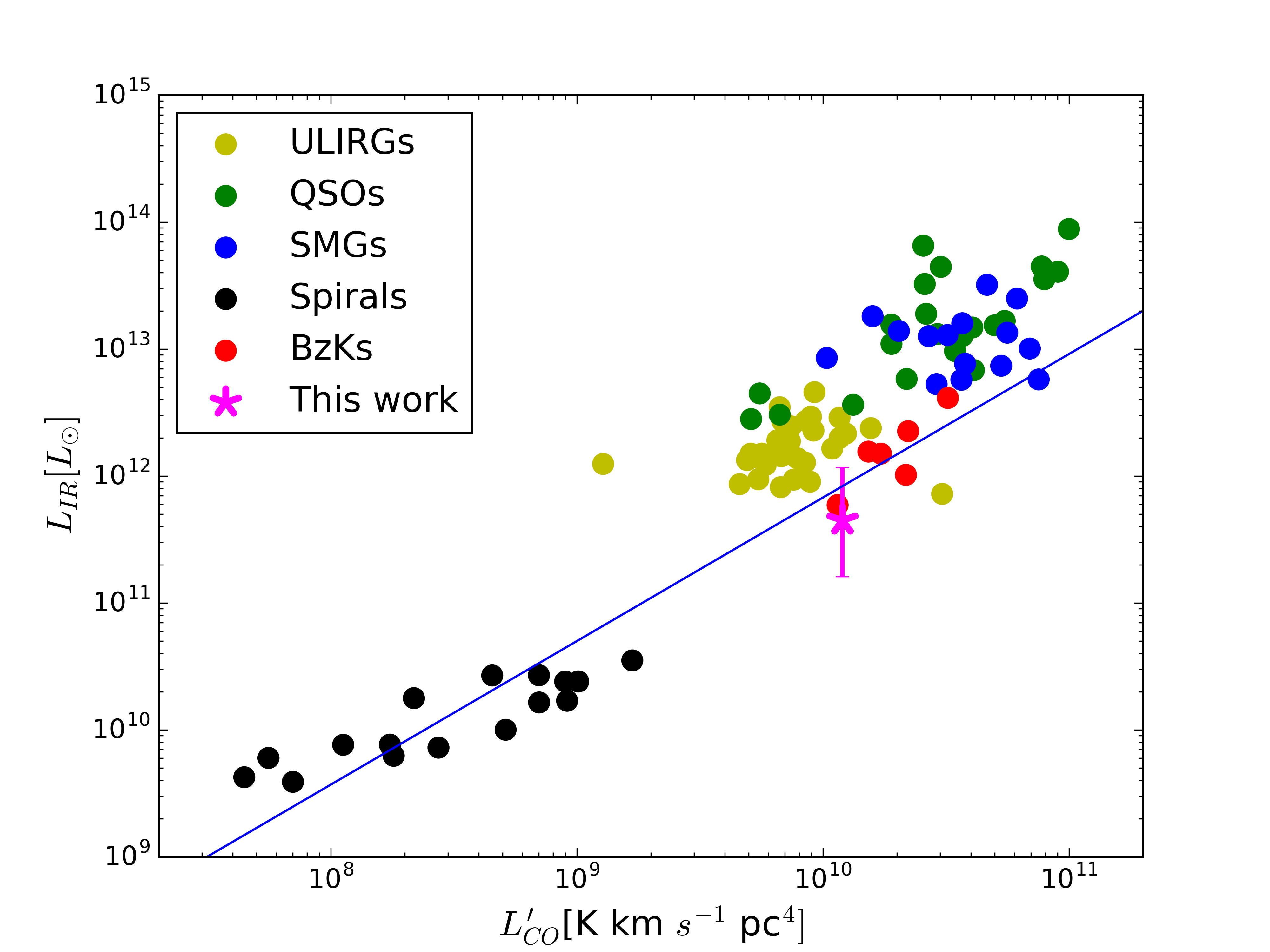}
\includegraphics[height=6cm,angle=0]{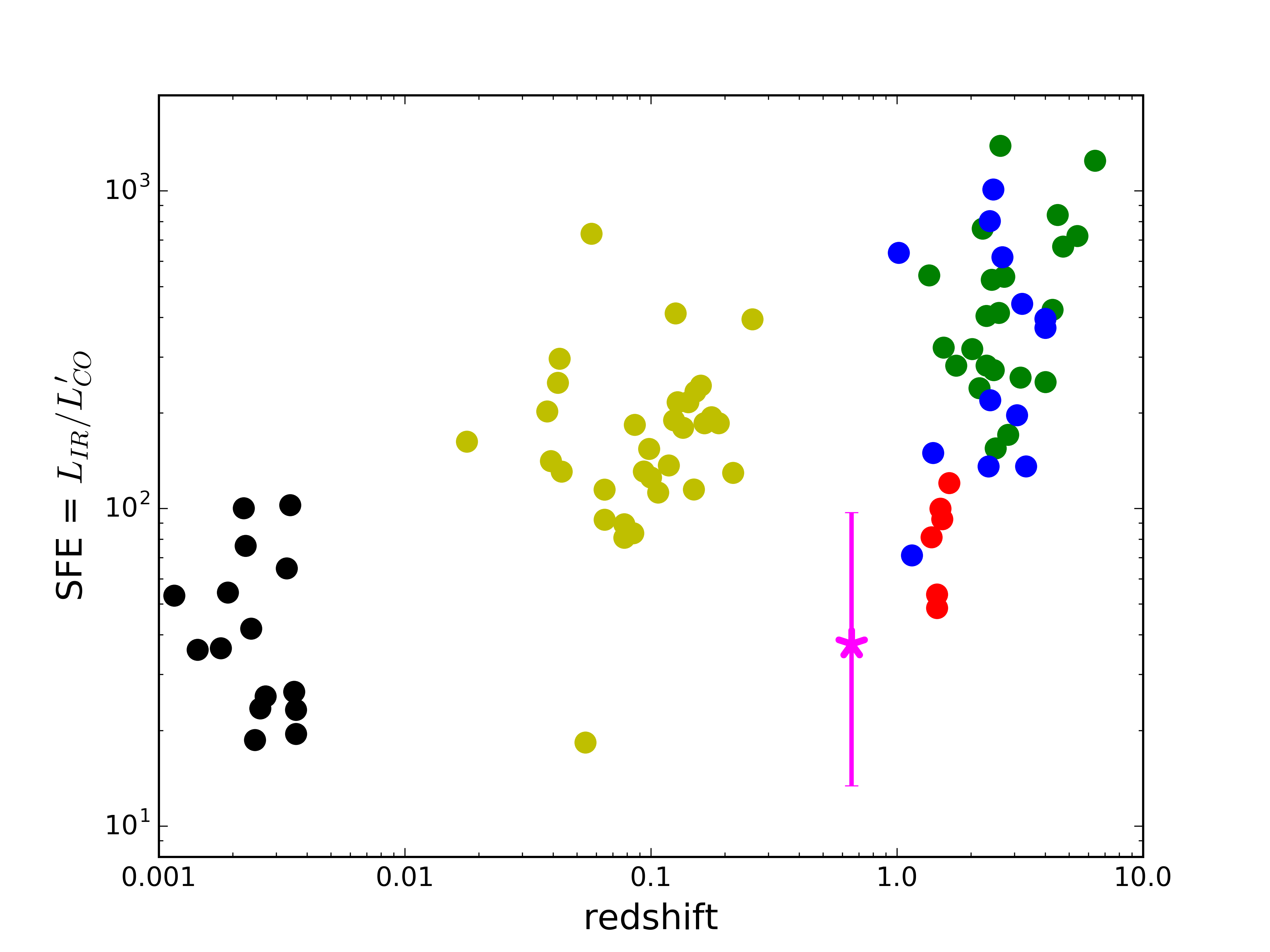}\\
\includegraphics[height=6cm,angle=0]{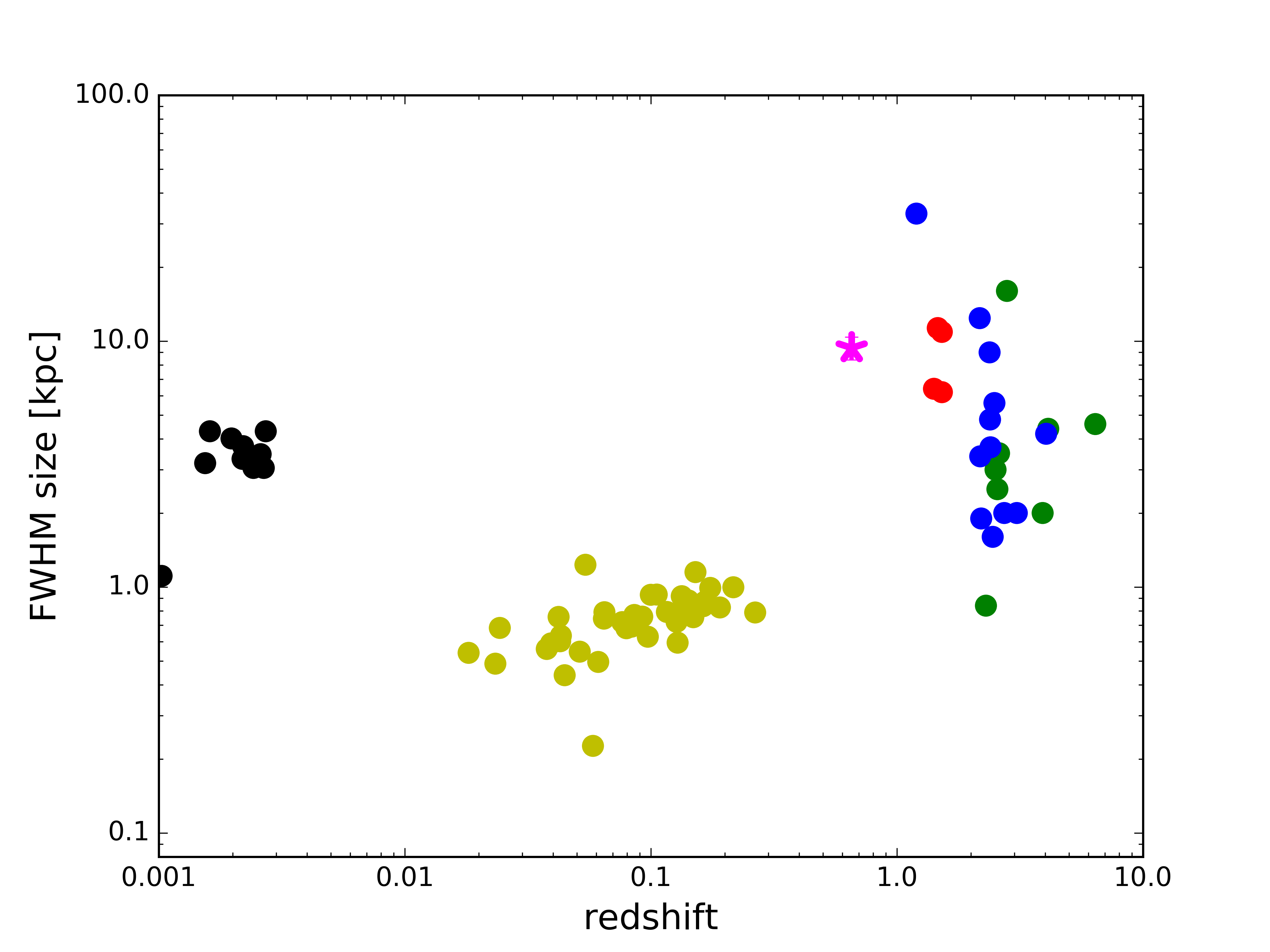}
\caption{Comparison of CO and IR properties of \obj\ to various local and distant samples as in \citet{2010ApJ...713..686D}. (Upper-left) CO vs IR luminosity. (Upper-right) star-formation efficiency (SFE) as a function of redshift. (Lower) FWHM CO region size as a function of redshift. Individual symbols indicate: SMGs \citep[blue circle,][]{2005MNRAS.359.1165G, 2009ApJ...695L.176D, 2008ApJ...680L..21F}, QSOs \citep[green circles,][]{2006ApJ...650..604R, 2005ARA&A..43..677S}, local ULIRGs \citep[yellow circles,][]{1997ApJ...478..144S}, local spirals \citep[black circles,][]{2008AJ....136.2782L, 2009ApJ...693.1736W}, BzK galaxies \citep[red circles,][]{2010ApJ...713..686D}. For local spirals, we associate a Hubble-flow redshift (right panel) based on the actual distance. The solid line in the upper left panel shows the best-fitting $L_{\rm CO}$ vs. $L_{\rm IR}$ relation to the combined sample of local spirals and distant BzK galaxies. The star-formation efficiency (SFEs) is defined as the ratio $L_{IR}$/$L'_{\rm CO}$. For the CO sizes, only resolved sources from individual samples have been used. Given the limited number of SMGs in \citet{2010ApJ...713..686D} with resolved CO emission, in the bottom panel, we include the SMG sizes reported by \citet{2008ApJ...680..246T}}
\label{fig:CO_properties_comparizon}
\end{center}
\end{figure*}

\subsection{Dynamical mass and CO--H$_2$ conversion factor}

In Fig.~\ref{fig:CO_recontructed_all} we also show the rotation velocity as a function of radial distance from the host galaxy centre (defined by the position of the AGN). To do so, we extract the velocity information along the major axis of the galaxy for each of the 8 velocity bins (84~km\,s$^{-1}$ velocity resolution). The rotation curve is corrected for the inclination of the disk $i=54^{\circ}$, which we find from the reconstructed morphological axial ratio. While a full dynamical analysis of the disk (joint with the continuum) is out of the scope of this work, we estimate the total enclosed dynamical mass, assuming the gas to be virialized, using
\begin{equation}
M_{\rm dyn}= \frac{\sigma^2R}{G},
\end{equation}
where $\sigma$ is rotational velocity ($\sigma=V_{\rm rot}/\sin i$) at radius $R$ and $G$ is the gravitational constant. Our inclination-corrected mass within 5 kpc is $M(r<5~{\rm kpc})=1.46\pm0.31 \times 10^{11}$~M$_\odot$.  

We also derive the Toomre parameter, according to the standard instability analysis defined by \citet{1964ApJ...139.1217T}, using 
\begin{equation}
Q_g=\frac{c_s\,k}{\pi G \Sigma},
\end{equation}
where $c_s$ is the sound speed that we approximate with mean velocity dispersion, $k$ is epicyclic frequency $k=\sqrt{2}V_{rot}/R$, $G$ is the gravitational constant and $\Sigma$ is the surface density of the disk at radius $R$. We find from our high angular resolution ALMA data that $Q = 1.07$ for the \obj\ molecular gas distribution, consistent with a stable rotating disk for the host galaxy.

From our dynamical mass, we estimate the gas mass by assuming a contribution of 30\% from the dark matter halo and a stellar mass of $M^* = 3.0\pm1.0 \times 10^{10}$~M$_{\odot}$, the latter derived by \citet{2017ApJ...836..180L} from the magnification corrected optical/infrared spectral energy distribution. We find a gas mass of $M_{H_2} = 8.3\pm3.0 \times 10^{10}$~M$_{\odot}$, which implies a CO--H$_2$ conversion factor of $\alpha = 5.5\pm2.0$~M$_{\odot}$\,(K~km\,s$^{-1}$~pc$^2$)$^{-1}$ for \obj.

\subsection{Comparison with far-infrared bright galaxy populations}

In Fig.~\ref{fig:CO_properties_comparizon} we show the $L'_{\rm CO}$ and $L_{IR}$ luminosity of \obj\ in comparison, as in \citep{2010ApJ...713..686D}, with various local and distant samples of dusty star-forming galaxies, both with and without evidence of quasar emission, and the star-formation efficiency ($L'_{\rm CO}$ / $L_{\rm IR}$) for the sample of galaxies as a function of redshift. In addition, the last panel of Fig.~\ref{fig:CO_properties_comparizon} shows the comparison of the CO emission region sizes. 

\obj\ has very similar properties to both local spiral galaxies and $z=1-2$ BzK galaxies \citep{2010ApJ...713..686D}. First, the $L'_{\rm CO}$ in \obj\ is directly comparable to that of high-redshift BzK galaxies. While the average SFE of BzKs is 1.5 times larger than that of \obj, this discrepancy might be due to the limited coverage of the FIR part of the spectral energy distribution of \obj.

Finally, the size of the molecular gas reservoir in \obj\ as traced by the CO (2-1) line agrees well with that  of BzK galaxies ($\sim$10~kpc FWHM), while being only slightly larger than in local spirals.

On the other hand, we clearly see that the molecular gas and heated dust properties of \obj\ are notably different from the population of ultra-luminous infrared galaxies (ULIRGs), sub-mm galaxies (SMGs) and high-redshift quasars. 

In particular, the large extent of the CO (2-1) emission stands in contrast to previous studies of molecular disks around $z_s$=1-4 gravitationally lensed quasars \citep{2005ARA&A..43..677S, 2011ApJ...739L..32R, 2016ApJ...827...18S}, which have a typical diameter of 1-5~kpc indicating that the molecular gas is concentrated in a compact circumnuclear region. No such circumnuclear concentration of molecular gas is seen in CO (2-1) in Fig. \ref{fig:CO_recontructed_all}, rather, the extended CO (2-1) emission makes \obj\ morphologically more similar to disk galaxies, rather than high-redshift quasars. However, we can not rule out a presence of a higher-excitation CO reservoir in the central region: future higher $J$-level imaging of CO at high angular resolution are required to confirm an existence of such a reservoir.

This suggests that the global properties of these galaxies are not determined by their AGN activity, but by their host galaxy morphology. This is also seen in our estimate of the CO--H$_2$ conversion factor, where for \obj\ the value is consistent with typical disk galaxies both at low redshift (e.g. the Milky Way) as well as higher redshifts ($z\sim 1.5$; \citealt{2010ApJ...713..686D}).

\section{Conclusions}
\label{concl}

We have presented new high-resolution ALMA continuum and CO (2-1) spectral line imaging of the star-forming and AGN composite galaxy \obj. Our analysis illustrates the combined power of gravitational lensing and high resolution interferometry as a tool to study distant galaxies and the co-evolution of AGN activity with the host galaxy star-formation properties. Our spectral line imaging, after correcting for the distortions from the gravitational lensing, represent the highest angular resolution mapping of CO (2-1) in an AGN system beyond the local Universe (see also \citealt{2008ApJ...686..851R}). We find that the CO (2-1) molecular gas distribution is extended on the order of $\sim 21$~kpc across the host galaxy, with a well-ordered velocity field consistent with a rotating disk, but with significant structure both in the molecular gas intensity and velocity dispersion, that we attribute to ongoing star-formation within the quasar host galaxy. 

Although the optical and mm-wave emission of \obj\ is dominated by its AGN, its star-formation efficiency, the size of the molecular disk and CO--H$_2$ conversion factor make it more similar to the nearby and high-redshift disk galaxies, rather than other high-redshift quasars. This suggests that the star-forming properties of these systems are dependent on the morphology of the host galaxy, as opposed to the level of AGN activity. This is consistent with the similarity in the molecular gas and dust luminosities of high redshift dusty star-forming galaxies and AGN \citep{2016ApJ...827...18S}. In particular, our finding that the CO--H$_2$ factor is similar to the Milky-Way derived value can have significant consequences for the applicability of scaling relations to estimate the molecular gas masses of high-redshift infrared-bright AGN systems. We note that such a large CO--H$_2$ factor has also been reported for the very high-redshift ($z \sim 4$) gravitationally lensed quasar APM~08279+5255 \citep{2007A&A...467..955W}.

\begin{acknowledgements}
This paper makes use of the following ALMA data: ADS/JAO.ALMA 2013.1.01207.S. ALMA is a partnership of ESO (representing its member states), NSF (USA) and NINS (Japan), together with NRC (Canada), NSC and ASIAA (Taiwan), and KASI (Republic of Korea), in cooperation with the Republic of Chile. The Joint ALMA Observatory is operated by ESO, AUI/NRAO and NAOJ. S. H. Suyu thanks the Max Planck Society for support through the Max Planck Research Group. This research is supported by the Swiss National Science Foundation (SNSF).
\end{acknowledgements} 

\bibliographystyle{aa}
\bibliography{paraficz}
\end{document}